\documentclass[twocolumn]{aastex62}
\usepackage{graphicx}
\received{2020 May 6}
\accepted{2020 June 17}
\submitjournal{AJ}
\shortauthors{Curiel et al.}

\begin{document} 

\title{An astrometric planetary companion candidate to the M9 Dwarf TVLM 513$-$46546}

\correspondingauthor{Salvador Curiel}
\email{scuriel@astro.unam.mx}

\author[0000-0003-4576-0436]{Salvador Curiel}
\affil{
Instituto de Astronom{\'\i}a,
Universidad Nacional Aut\'onoma de M\'exico (UNAM),
Apdo Postal 70-264,
M\'exico, D.F., M\'exico.
}

\author[0000-0002-2863-676X]{Gisela N. Ortiz-Le\'on}
\affiliation{Max Planck Institut f\"ur Radioastronomie,
Auf dem H\"ugel 69, D-53121 Bonn, Germany}

\author{Amy J. Mioduszewski}
\affiliation{National Radio Astronomy Observatory,
Domenici Science Operations Center,
1003 Lopezville Road, Socorro,
NM 87801, USA }

\author[0000-0002-7179-6427]{Rosa M. Torres}
\affiliation{Centro Universitario de Tonal\'a, Universidad de Guadalajara,
Avenida Nuevo Perif\'erico No. 555, Ejido San Jos\'e Tatepozco, 
C.P. 48525, Tonal\'a, Jalisco, 
M\'exico }
 
\begin{abstract}

Astrometric observations of the M9 dwarf TVLM 513--46546 taken with the VLBA reveal an astrometric signature consistent with a period of 221 $\pm$ 5 days. The orbital fit implies that the companion has a mass m$_{p}$ = 0.35$-$0.42 $M_{J}$, a circular orbit ($e \simeq 0$), a semi-major axis  a = 0.28$-$0.31 AU and an inclination angle  i = 71$-$88$^\circ$. The detected companion, TVLM~513$b$, is one of the few giant-mass planets found associated to UCDs. The presence of a Saturn-like planet on a circular orbit, 0.3 AU from a 0.06$-$0.08 $M_\odot$ star,
represents a challenge to planet formation theory. 
This is the first astrometric detection of a planet at radio wavelengths.

 \end{abstract}

\keywords{astrometry $-$ circumstellar matter $-$ planetary systems $-$
                   stars: coronae $-$ stars: individual (TVLM 513$-$46546) $-$ stars:
                  X-rays: stars
               }

%

\section{Introduction} \label{sec:intro}

The search for extrasolar planets is one of the most vibrant fields in modern astrophysics. Thanks to the advent of new instrumentation and data analysis methods, many exoplanets have been discovered and characterized in recent years. Currently, one of the main targets for exoplanet  searches are main-sequence low-mass stars, known as M dwarfs, since they are the most numerous stars in the Galaxy and are known to host a large number of small planets \citep[e.g.,][]{chabrier00, bonfils13, dressing13, gillon17}.  However, the occurrence of giant-mass planets around M dwarfs is low compared to their occurrence around Sun-like stars \citep[e.g.,][]{endl06, cumming08, bonfils13}, which is consistent with core-accretion models that predict few Jovian-mass planets orbiting M dwarfs \citep[e.g.,][]{laughlin04, adam05, ida05, kennedy08}.

Ultracool Dwarfs (UCDs) are at the lower mass end of the M dwarf stellar class. The occurrence of giant planets around UCDs is an important observational constraint for planet formation theories. For instance, the core-accretion theory predicts that giant-mass planet formation scales with the central star mass; therefore, giant-mass planet formation is expected to be low around M dwarfs and, especially, around UCDs  \citep[e.g.,][]{laughlin04, kennedy08}. Giant planets around UCDs can also be formed via disk instability if their disks are suitably unstable \citep[e.g.,][]{boss06}.

In recent years, UCDs have been found to host Earth- and Mars-mass planets  \citep[e.g.,][]{kubas12, muirhead12, gillon17}. Radial velocity (RV) measurements of this kind of star have been used to search for giant planets on compact orbits, excluding a large population of giant-mass planets on very tight orbits $<$0.05 AU \citep[e.g.,][]{blake10, rodler12}. Direct imaging searches, which are adequate to search for giant-mass planets with wide orbits, exclude a large population of gaseous planets at wider separations than 2 au \citep[e.g.,][]{stumpf10}. 

In the near future, Gaia's astrometric observations have the potential to detect many (probably thousands) exoplanets and brown dwarfs associated with solar-type and low-mass stars \citep[e.g.,][]{casertano08, sozzetti14, perryman14}. Very long baseline interferometry (VLBI) astrometric observations can also reveal sub stellar companions (brown dwarfs and giant-mass planets) around pre-main-sequence stars and M-dwarf stars. This technique has already yielded mass upper limits of a few planetary companion candidates \citep[][]{bower09, bower11}. Observations carried out in the optical wavelength range with 10 m class telescopes allow similar astrometric searches, but they require conversion of relative to absolute astrometry \citep[e.g.,][]{sahlmann16}.

Astrometric planet searches consist of measuring the positional shift (or reflex motion) of the star around the center of mass of the orbit due to the gravitational pull of a companion. This technique allows the discovery and characterization of extrasolar planets, provided that the astrometric accuracy is much smaller than the amplitude of the reflex motion. For instance, a reflex motion of 1 mas will be produced by a 5 Jupiter-mass planet on a three-year orbit around a Sun-like star at 10 pc \citep[][]{sozzetti05, sahlmann12}.
Several astrometric planet searches have been conducted toward UCDs, but they have not yet found  new exoplanets \citep[e.g.,][]{pravdo96, boss09, forbrich13}. However, these searches have been crucial primarily because they have enabled the determination of precise trigonometric distances, which are important to determine the luminosity, mass, and ages of UCDs. These properties are central to understanding the physics of these objects  \citep[e.g.,][]{dahn02, andrei11, dupuy12, dupuy13, smart13, sahlmann14}.

At present, low-mass brown dwarfs (several tens of Jupiter masses) have been found orbiting a few UCDs and TTauri stars \citep[e.g.,][]{sahlmann13, curiel19}.
Until now, only a few UCDs have been studied with VLBI 
(TVLM 513$-$46546, \cite{forbrich13}; LSPM J1314+1320AB, \cite{dupuy16}; \cite{forbrich16}). In
LSPM J1314+1320AB, a close binary system, only one of the sources is detected at radio wavelengths. On the other hand, the M9 UCD dwarf TVLM 513$-$46546 \citep[hereafter TVLM~513,][]{reid08, west11} is an apparent single star that was first detected at radio wavelengths by \cite{berger02}, who found persistent emission and a circularly polarized flare lasting about 15 minutes at 8.5 GHz. Astrometric VLBI monitoring has yielded a precise trigonometric parallax implying a distance of 10.762$\pm$0.027 pc \citep{forbrich13, gawronski17}.
The bolometric luminosity and a lack of Li absorption lines imply a minimum age of $\sim$400 Myr and a mass between 0.06 and 0.08 $M_\odot$ \citep[][]{martin94, reid02, hallinan08}, while membership in the “young/old disk” kinematic category of 
\cite{leggett92} is suggested by a low space velocity \citep{leggett98}. This estimated mass places TVLM~513 just at the brown dwarf boundary \citep{hallinan06}.
\citet{forbrich13} investigated the possibility that the residuals of their parallax fit could be associated with the reflex motion of the M dwarf due to an unseen companion. In their analysis, they considered only circular orbits on the plane of the sky. 
However, even when the residuals are significantly larger than the astrometric precision, their analysis suggested that the VLBI astrometry, in principle, excludes the presence of unseen companions with masses higher than $\sim$4 M$_{J}$ at orbital periods of $\sim$10 days or $\sim$0.3 M$_{J}$ at periods $\sim$710 days \citep{forbrich13}. \citet{gawronski17} also excluded the possibility of companions more massive than Jupiter in orbits with periods longer than $\sim$1 yr.
In addition, near-IR imaging excludes companions with separations between 0.1 and 15 arcsec 
\citep{close03}.

In this paper, we investigate the possibility that the reflex motion due to a companion orbiting TVLM~513 is the responsible for the relatively large residuals of the astrometric fit of multi-epoch observations of its nonthermal radio emission. We present new Very Long Baseline Array (VLBA) observations of this source taken over a period of about 1.5 yr. We have also recalibrated previous VLBA observations, which we combine with the new data to search for evidence of a putative companion around this UCD. We describe our observations and calibration strategy in Section 2. In Section 3 we present the fitting procedure. In section 4 we present the results and the discussion, and the main conclusions are presented in Section 5.


\section{Observations} \label{sec:obs}

We use the VLBA to conduct new observations of TVLM~513 over a time interval of 1.5 yr starting in   2018 June. A total of 18 epochs were observed as part of projects BC236, BC244, and BC255 (see Table 1 for details). The observations were taken at a frequency of 8.4~GHz in dual polarization mode with 256 MHz or 512 MHz (last two epochs) of total bandwidth in each polarization. The observations consisted of alternate scans on the target and the phase-reference calibrator, J1455+2131, with an on-source time of $\sim$1 minute each. Scans on the calibrators J1513+2338, J1453+2648, and J1511+2208 were observed every $\sim$50 minutes to improve the phase calibration. In addition, two geodetic-like blocks of  $\sim$30 minutes each were included at the beginning and the end of the observing sessions. These scans are used to estimate and remove phase offsets introduced by tropospheric and clock errors. 

To complement our analysis, we also include in this paper archival VLBA data from project BF100, which used the same phase calibrator and observed in a total of nine epochs  (see also Table~\ref{tab_1}) from 2010 March to 2011 August. These data were taken at a frequency of 8.4~GHz with 64~MHz of total bandwidth in dual polarization mode. 

We use AIPS \citep{greisen03} to reduce our new and the archival data following standard procedures for phase-referencing observations \citep[e.g.,][]{ortizleon17, curiel19}. 
Particular care was taken when calibrating the archival data since they used an old position for the phase calibrator during correlation. Then, before deriving any calibration, we correct the position of the phase calibrator to the new position as measured in our new data. Offsets of $-$0.10~mas in right ascension and $+$0.39~mas in declination were added in the first 7 epochs of BF100. The calibrator position was updated in the last two epochs, resulting in offsets of $+$0.01 and $-$0.01~mas in R.A.  and decl., respectively.

\begin{figure*}
 \begin{center}
   \plotone{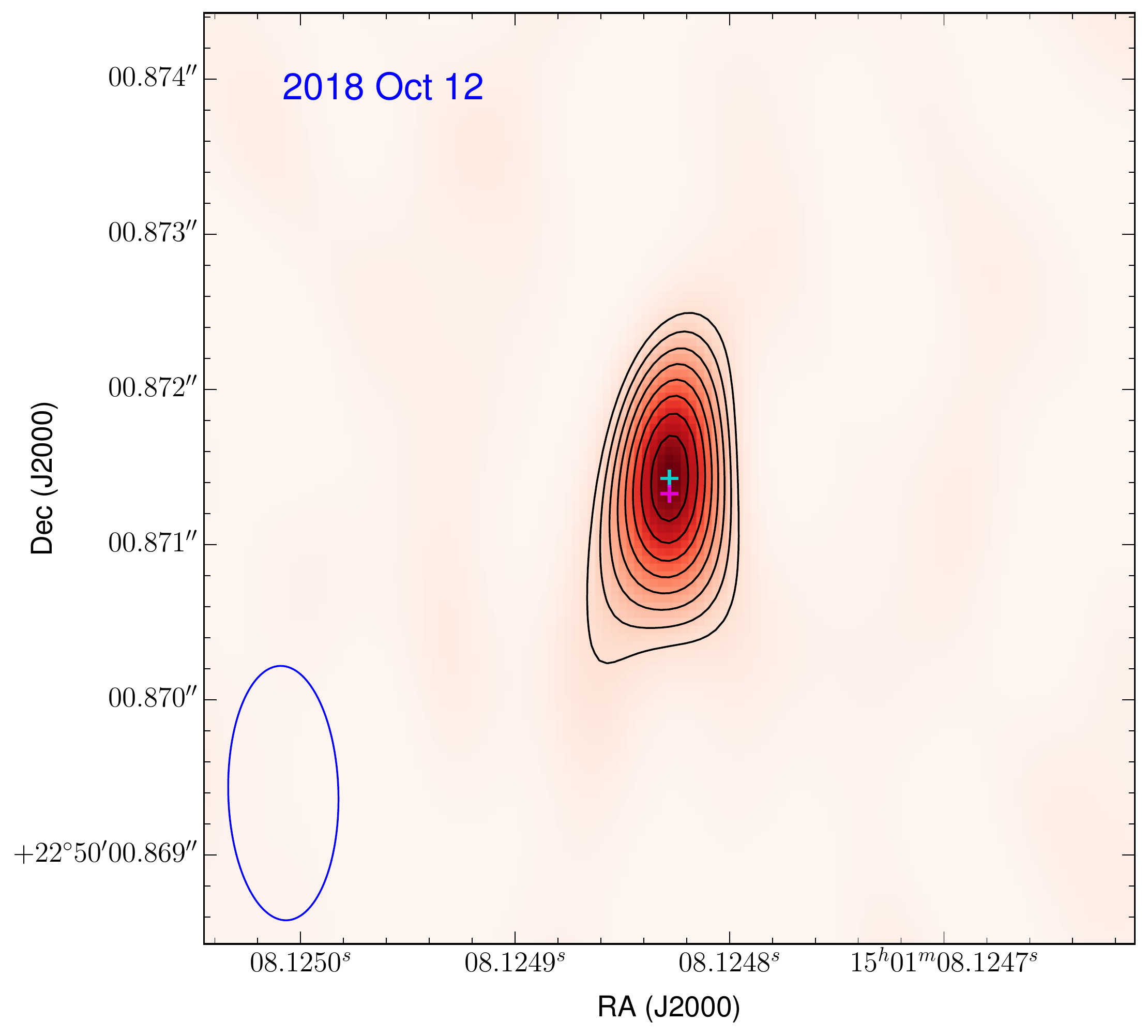}
 \end{center}
  \caption{The intensity map of TVLM~513 taken on 2018 October 12 is shown here as an example. The contours are 4, 5, and 6$\times\sigma$, where $\sigma=19~\mu{Jy~beam}^{-1}$ is the rms noise level. The plus signs mark the fitted peak positions obtained with the maximum-fit algorithm MAXFIT (cyan) and with a Gaussian brightness distribution obtained with JMFIT (magenta). }
  \label{fig_1}%
\end{figure*}

The calibrated data were imaged within AIPS using a pixel size of 50~$\mu$as and two weightings schemes, pure natural (robust parameter = 5) and partial uniform (robust = 0). Detections of TVLM~513 were achieved in all 18 new epochs, and six epochs of the old project BF100. Our images have, on average, rms noise levels of $\sim14~\mu$Jy~beam$^{-1}$ for natural weighting, i.e.\ three times better than previous VLBA observations \citep{forbrich13}, as a result of increased bandwidth and larger integration time. 
Source positions and positional uncertainties for pure natural and partial uniform weighting were first obtained by fitting a Gaussian model to the source brightness distribution. This was done using the AIPS task JMFIT. In addition, we measured the position of the pixel with maximum flux density using the task MAXFIT. Table~\ref{tab_1} gives the measured positions with MAXFIT for partial uniform weighting. 
To estimate the errors in positions, we use the equation for the expected theoretical astrometric uncertainty given by

\begin{equation}
\sigma \simeq \frac{\theta_{\rm res}}{2 ~{\rm S/N}},
\end{equation}

\noindent
where $\theta_{\rm res}$ is the resolution of the interferometer, and S/N the signal-to-noise ratio \citep{thompson17}. Then we quadratically added half of the pixel size to this uncertainty. For each epoch, the resolution was taken as the geometric mean of the major and minor size of the telescope beam. The S/N is directly provided by JMFIT. 
Figure~\ref{fig_1} shows the intensity map obtained on 2018 October 12 with partial uniform weighting. Also shown are the positions as measured with JMFIT and MAXFIT. We see that the JMFIT position does not coincide with the position of the pixel with maximum flux. This is because the procedure used to obtain the centroid is affected by the asymmetry of the emission. Here JMFIT is also sensitive to the box selected to define the region in the image to be fitted. The MAXFIT positions are not affected by source asymmetries, therefore, they provide a better estimation of the star position.

\section{Fitting of the Astrometric Data}

\subsection{Least-squares Periodogram} \label{sec:rlscp}

We use a periodogram code to search for astrometric signatures that indicate the possible presence of one or more companions to the main source. The periodogram of the astrometric data is obtained using a modified version of the classic least-squares periodogram method described by \cite{curiel19}. 
The new version of the code,  which we call a recursive least-squares periodogram with a circular orbit (RLSCP), takes into account the possibility of fitting the Keplerian orbits of several companions \citep[see also, e.g.,][]{anglada12}. 
This recursive periodogram consists of fitting all the parameters of the already detected signals together with the signal of a new companion, which is under investigation. We start assuming circular orbits, but we can include a fixed eccentricity for the signals already found. When no previous planets have been detected, the initial periodogram is obtained by comparing the least-squares fits of the basic model (proper motions and parallax only) and a one-companion model (proper motions, parallax, and Keplerian orbit of a single companion). When a signal has already been detected, the recursive periodogram compares  the least-squares fits of a one- and two-companion model (proper motions, parallax, and Keplerian orbits of two companions), and so on. 

The weighted least-squares solution is obtained by fitting all of the free parameters in the model for a given period. The sum of the weighted residuals divided by N$_{obs}$ is the so-called $\chi^{2}$ statistic, where N$_{obs}$ is the number of data points. Then, each $\chi^{2}_{P}$ of a given model with $k_{P}$$-$free parameters can be compared to the $\chi^{2}_{0}$ of the null hypothesis with $k_{0}$$-$free parameters by computing the power, $z$, as \citep[e.g.,][]{anglada12, curiel19}:

\begin{equation}
z(P)  =  \frac{(\chi^{2}_{k} - \chi^{2}_{P})/(N_{k+1} - N_{k})}{\chi^{2}_{P}/(N_{obs} - N_{k+1})}, 
\end{equation}

\noindent
where $\chi^{2}_{k}$ is the $\chi^{2}$ statistic for the model with $k$ planets (the null hypothesis), $\chi^{2}_{P}$ is the $\chi^{2}$ statistic for the model including one more planet with an orbital period $P$, $N_{k}$ is the number of free parameters in the model with $k$ planets, and  $N_{k+1}$ is the number of free parameters in the model including one more candidate in a circular orbit with an orbital period $P$.
In this model, a large $z$ is interpreted as a very significant solution. The values of $z$ follow a Fisher $F$-distribution with $N_{k+1} - N_{k}$ and $N_{obs} - N_{k+1}$ degrees of freedom
\citep{scargle82, cumming04}. Even if only noise is present, a periodogram will contain several peaks \citep[see][as an example]{scargle82} whose existence has to be considered in obtaining the probability that a peak in the periodogram has a power higher than $z(P)$ by chance, which is the so-called false-alarm probability (FAP),

\begin{equation}
\rm{FAP} = 1 - (1 - \rm{Prob}[z > z(P)])^{O},
\end{equation}

\noindent
where $O$ is the number of independent frequencies. In the case of uneven sampling, $O$ can be quite large and is roughly the number of periodogram peaks one could expect from a data set with only Gaussian noise and the same cadence as the real observations. We adopt the recipe $O \sim \Delta T/P_{min}$ given in \citet[Section 2.2]{cumming04}, where $\Delta T$ is the time span of the observations and $P_{min}$ is the minimum period searched. For instance, assuming that $\Delta$$T$ = 560 days and $P_{min}$ = 20 days, the astrometric data is expected to have  $O \sim 28$  peaks.

\subsection{Least-squares and AGA Fitting Algorithms} \label{sec:ls-aga}

Here we use the least-squares algorithm and the asexual genetic algorithm (AGA) presented by \cite{curiel19}. In short, we use the source barycentric two-dimensional position described as a function of time ($\alpha(t)$, $\delta(t)$), accounting for the (secular) effects of proper motions ($\mu_\alpha$ and $\mu_\delta$), the (periodic) effect of the parallax $\Pi$, and the (Keplerian) gravitational perturbation induced on the host star by one or more companions, such as low-mass stars, substellar companions, or planets (mutual interactions between companions are not taken into account).
Given a discrete set of $N_{obs}$ data points ($\alpha(i)$, $\delta(i)$) with associated measurement errors $\sigma_{i}$, one seeks for the best possible model (in other words, the closest fit) for these data using a specific form of the fitting function, ($\alpha(t)$, $\delta(t)$). This function has, in general, several adjustable parameters, whose values are obtained by minimizing a $"$merit function,$"$ which measures the agreement between the data ($\alpha(i)$, $\delta(i)$) and the model function ($\alpha(t)$, $\delta(t)$). The maximum-likelihood estimate of the model parameters ($c_{i}$, ..., $c_{k}$) is obtained by minimizing the $\chi^{2}$ function \citep[e.g.,][]{canto09, curiel19}:

\begin{equation}
\begin{array}{l}
\chi^{2}_{min}  = {\sum_{i=1}^{N} \left( \frac{\alpha_{i} - \alpha(t_{i}; c_{1}, ..., c_{k})}{\sigma_{i}}\right)^{2}} \\
 ~~~~~~~  + {\sum_{i=1}^{N} \left( \frac{\delta_{i} - \delta(t_{i}; c_{1}, ..., c_{k})}{\sigma_{i}}\right)^{2}},   
\end{array}
\end{equation}


\noindent
where each data point ($\alpha_{i},\delta_{i}$) has a measurement error that is independently random and distributed as a normal distribution about the $"$true$"$ model with standard deviation $\sigma_{i}$.

\section{Results and Discussion} \label{sec:results}

The new VLBA astrometric observations of the M9~dwarf TVLM~513 cover a time span of about 558 days, with an observational cadence that varies during all the time observed. Including previous VLBA observations of this source, 
the time span of the observations increases to about 3574 days. However, the observations were carried out in two time blocks, one of about 1 yr and the other of about 1.5 yr, separated by about 7 yr (see Table~\ref{tab_1}). 
The time span and cadence of the new and the combined data are adequate to fit the proper motions and the parallax of this source, as well as to search for substellar companions with orbital periods between a few days and more than 1 yr. Below, we use the recursive least-squares periodogram (see sec.~\ref{sec:rlscp}), and the least-squares and the AGA algorithms presented by \cite{curiel19} to fit the astrometric data of this source.

The observation taken on 2018 November 5 was carried out under bad weather conditions. Six stations experienced precipitation or high winds during a significant part of the experiment, and Maunakea experienced technical issues. As a result, the quality of the image and the astrometry was affected. We found that this epoch shows high residuals of the parallax fit in comparison with that seen in the rest of the observations that were taken under better weather conditions. Therefore, we do not include this epoch in our analysis. Hence, we use a total of 23 epochs in the analysis we present here. For the astrometric fits that we present here, we have used the astrometric position of the source obtained with partial uniform weighting using the task MAXFIT and errors from equation 1 (see sec.~\ref{sec:obs} for more details).

\subsection{Single-source Astrometry}

\begin{figure*}
   \centering
    \plotone{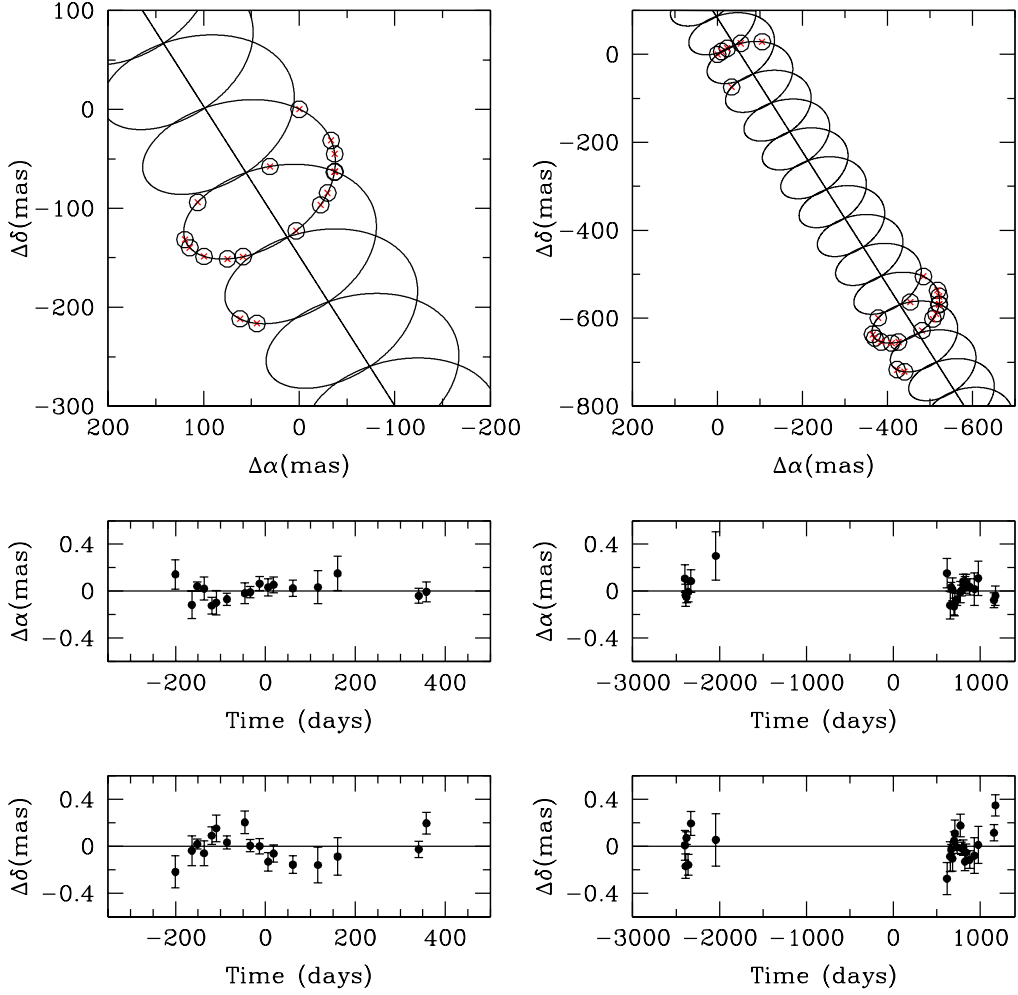}
    \caption{Parallax fit to the VLBA data. The left panels show the fit for the VLBA new data and the right panels show the fit of the combined VLBA data. The upper panels show the observed data and the astrometric fit obtained when fitting only the proper motions and the parallax of TVLM~513. The lower panels show the residuals in R.A. and decl. as a function of time. The residuals show a clear temporal trend that suggests that they could be due to at least one companion.}
    \label{fig_2}%
    \end{figure*}

First, we used both the least-squares and the AGA algorithms \citep{curiel19} to fit the proper motions and parallax to the 17 new VLBA astrometric observations without taking into account any possible companion (single-source solution). Then, we fitted all of the VLBA astrometric data, including six previous VLBA detections of this source, obtained by \cite{forbrich13} (see Table~\ref{tab_1}). 
The results of a single-source solution are shown in Table~\ref{tab_2} and Figure~\ref{fig_2}. We find that the fitted parameters (proper motions and parallax) are very similar in both cases. However, the residuals are large and show a temporal trend that suggests the presence of at least one companion with a possible orbital period of a few hundred days (see Figure~\ref{fig_2}).

We also fitted the astrometric data with acceleration terms, which take into account an astrometric signature due to a possible companion with a large orbital period. 
We find that the fits do not improve substantially when including acceleration terms (see Table~\ref{tab_3}). The fitted acceleration terms are small in the case of the combined VLBA data ($a_{\alpha}$ = $-$0.0144 $\pm$ 0.0045 mas~yr$^{-2}$ and $a_{\delta}$ = 0.0332 $\pm$ 0.0049 mas~yr$^{-2}$) and somewhat larger using only the new VLBA data, but they are consistent with zero within the errors. In this case the acceleration terms are $a_{\alpha}$ = $-$0.34 $\pm$ 0.30 and $a_{\delta}$ = 0.41 $\pm$ 0.33 mas~yr$^{-2}$, which suggests that this source might have a companion with an orbital period larger than about 1.5 yr (the time span of the new astrometric VLBA data) and smaller than 9.8 yr (the time span of  the combined astrometric VLBA data). 

In what follows, we obtain the astrometric fit of the data without taking into account possible acceleration terms.

\subsection{Single-companion Astrometry}

\begin{figure*}
   \centering
   \plottwo{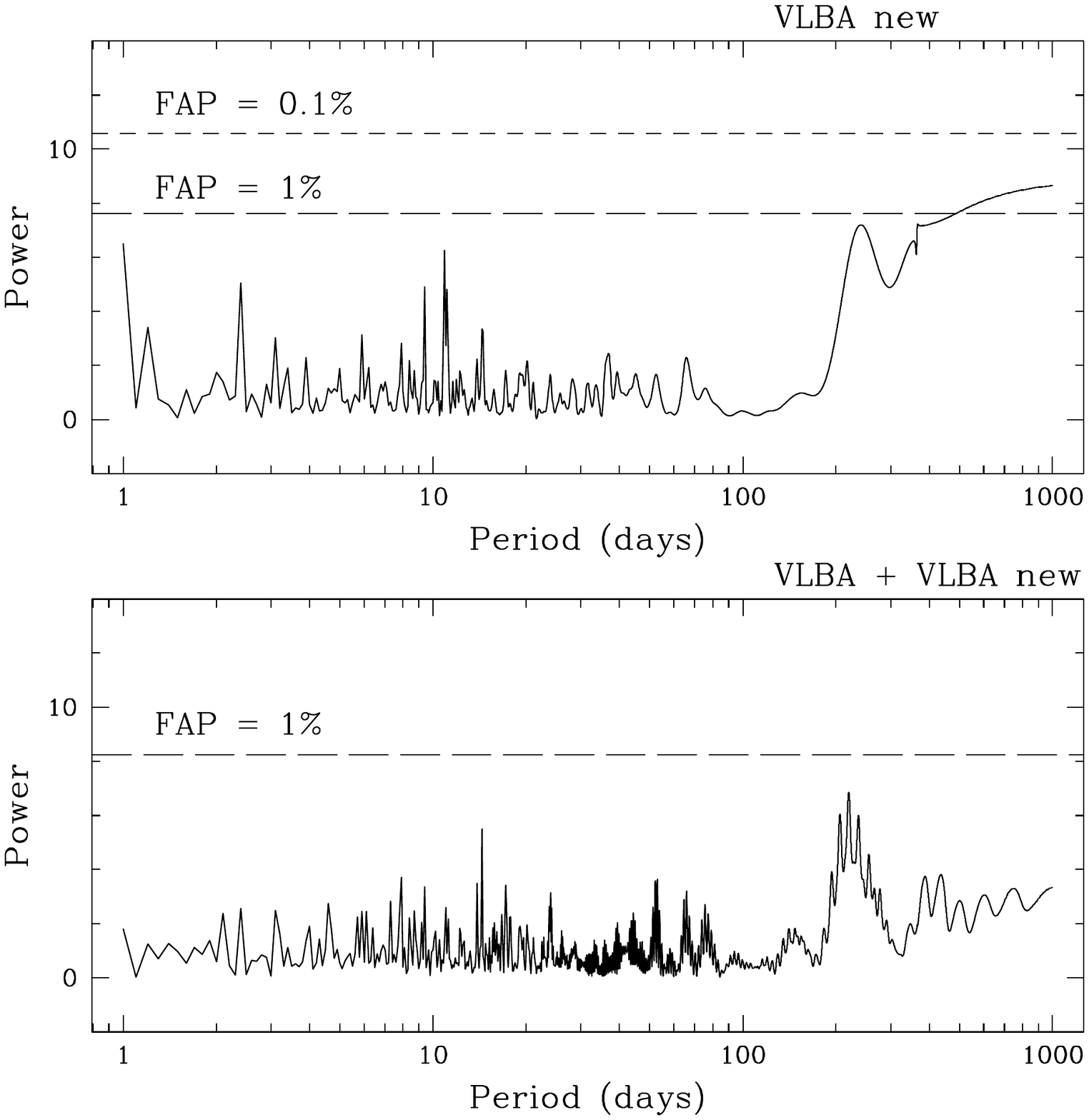}{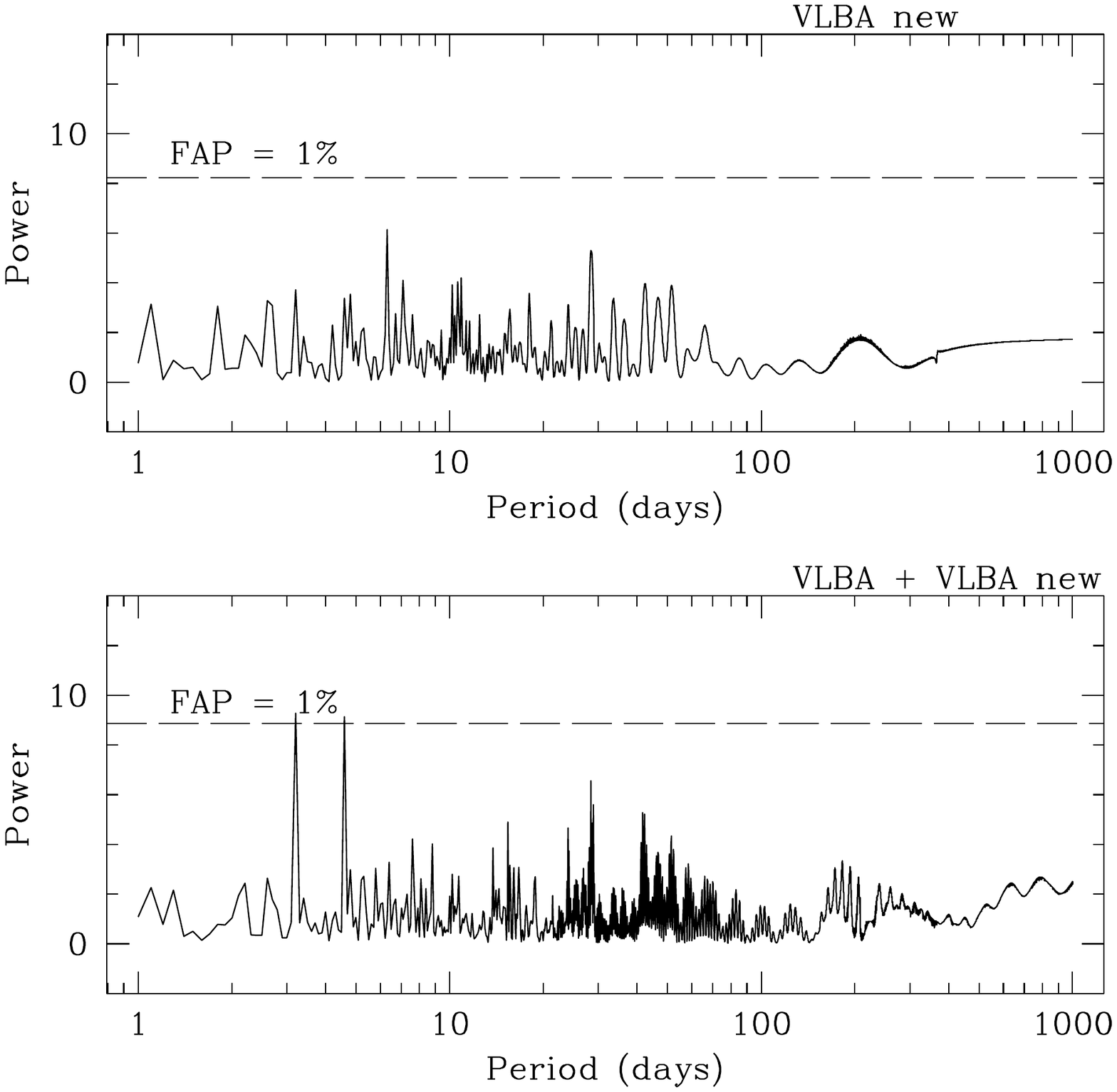}
    \caption{{\bf Left:} RLSCP periodogram obtained by fixing the eccentricity $e = 0$. The upper panel shows the periodogram obtained with the new astrometric VLBA data. The lower panel corresponds to the fit obtained with the combined VLBA data. The horizontal lines indicate the limits of the false alarm probabilities FAP = 1\% and 0.1\%. {\bf Right:} same as the left panels, but in this case the plot shows the RLSCP periodogram obtained by including two possible astrometric signals: the detected astrometric signal that appears in the initial periodogram (left panels) and the signal of a possible second companion. These periodograms do not show clear evidence of a second companion. However, the periodograms show a very weak temporal trend at orbital periods larger than 300 days, which may suggest the presence of a second companion. The two very narrow peaks between 3 and 5 days are most likely spurious signals or artifacts.}
    \label{fig_3}%
    \end{figure*}

The RLSCP of the new astrometric VLBA data (see Figure~\ref{fig_3}) does not show a narrow prominent peak. However, it shows a somewhat $''$broad signal$''$ with an orbital period between 200 and 300 days. The periodogram also shows that this broad signal is part of a large plateau-like structure that extends beyond the orbital periods considered in the plot (1000 days). This plateau has a drop in the periodogram power around 297 days, suggesting that there may be two broad signals in the periodogram, one at about 241 days and the other one with an orbital period larger than the time span of the new VLBA observations (about 1.5 yr). The main broad signal is not well constrained but seems to have a relatively weak peak at about 241 days. The FAP of this main peak is 1.43\%, suggesting that this signal is real and that it could be due to a companion. 

The RLSCP of the combined (old and new) VLBA data also shows a broad signal between 200 and 300 days, nearly coinciding with the broad signal observed in the periodogram of the new  data. In this case, the periodogram appears somewhat noisier, especially at orbital periods larger than 100 days. The main peak of the combined data is located at  220 days and has an FAP of 5.39\%, which, although it is somewhat above the 1\% limit usually used to consider a signal as possibly real, also suggests that the signal is real and due to the presence of a companion.

To further investigate the possibility that the peak that appears in the periodograms is real, we computed the recursive periodogram of the two data sets including the signal of this possible companion (two-companion solution). We now fit simultaneously the parameters of the already detected signal together with the signal under investigation (a second possible companion). To obtain an improved global solution (with two possible candidates), we include in the fitting the orbital period of the first companion using, as an initial guess, the  orbital period of the peak in the initial periodogram. The RLSCP algorithm includes the possibility that the orbital period of the first companion adjusts during the simultaneous fit of both possible companions. The resultant periodogram is shown in Figure~\ref{fig_3}. The new periodograms show that the signal of the initial candidate disappears, leaving some residual noise. In addition, the new periodograms show no significant signals, indicating that there is only one significant signal in the periodogram. The new periodogram of the combined data shows two very narrow and relatively strong signals between 3 and 5 days that do not appear in the periodogram of the new VLBA  data. These signals are most likely spurious signals or artifacts. The new periodograms also show a slow rising signal close to the end of the plot. This suggest that there may be a second companion that, if real, produces a small astrometric signal and that its orbital period is larger than the time span of the new VLBA data ($>$ 558 days). Further observations will be needed to confirm this putative second companion.

\begin{figure*}
   \centering
   \plottwo{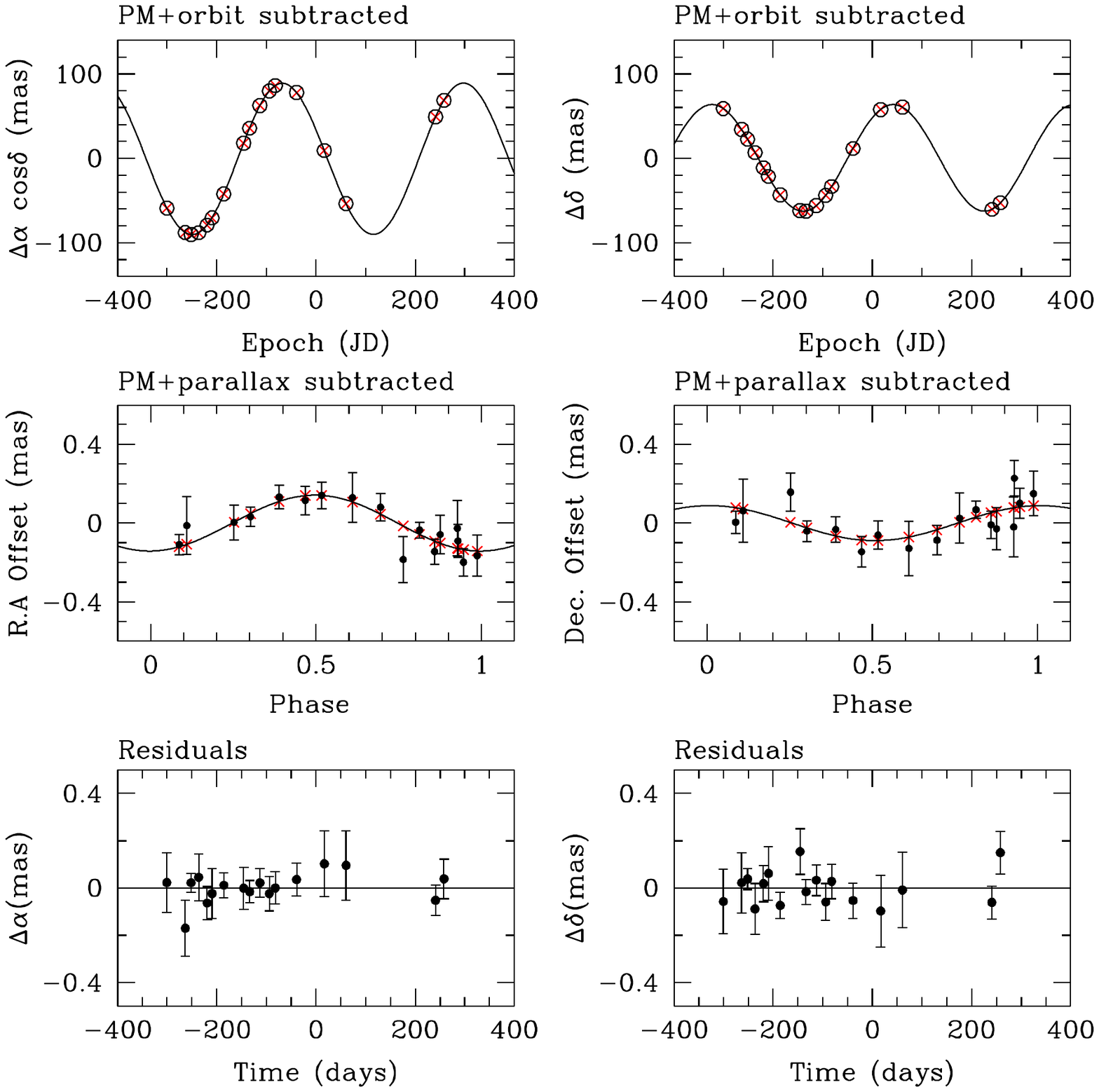}{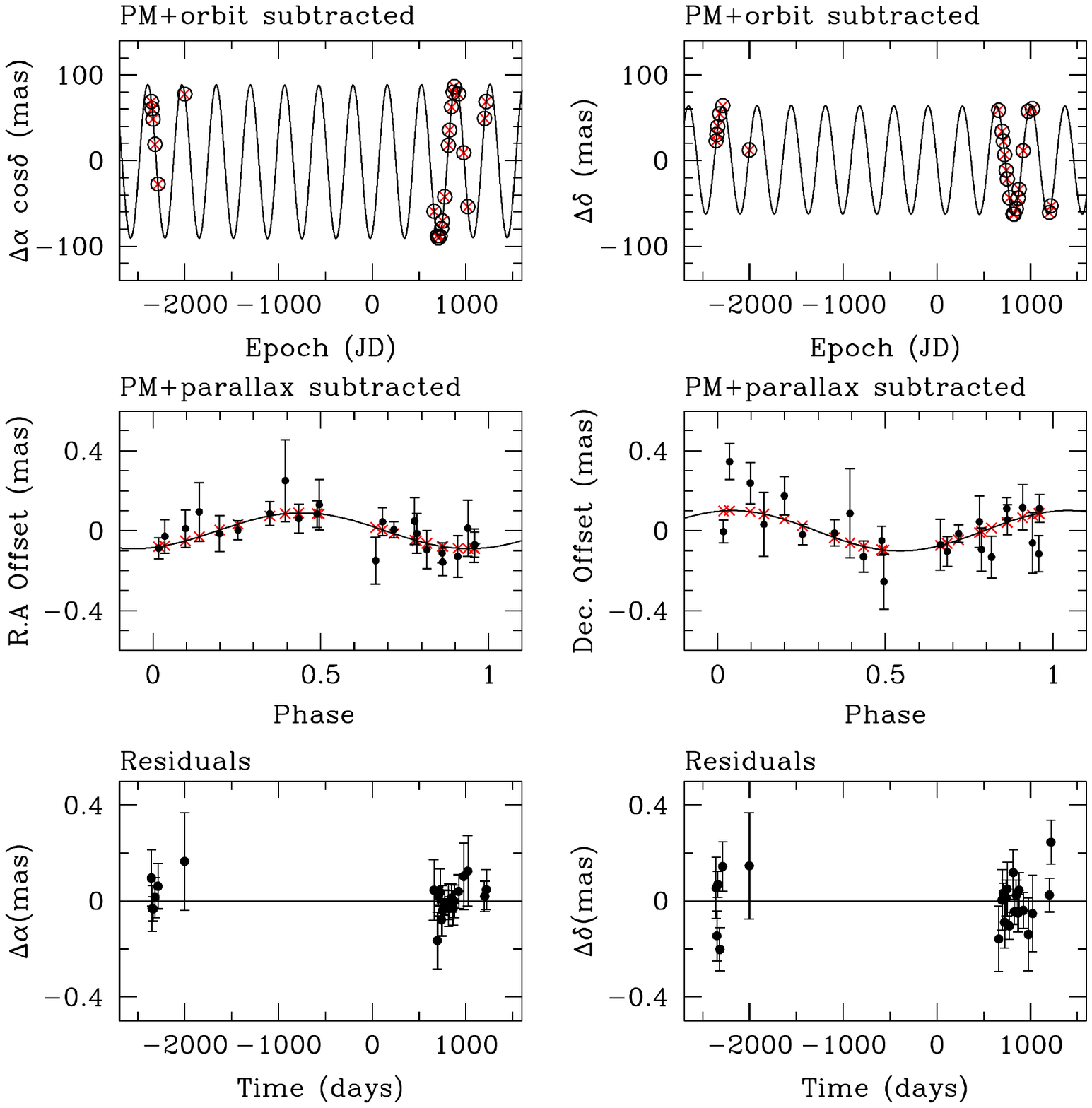}
    \caption{Single-companion astrometric fit of TVLM~513 using only the new (left) and the combined (right) VLBA data. The upper two panels show the parallax fit of the source after subtracting proper motions and the astrometric signal of the companion. The middle panels show the astrometric fit of the companion after removing parallax and proper motions. The lower panels show the residuals of the astrometric fit.}
    \label{fig_4}%
    \end{figure*}

We then used both the least-squares and the AGA algorithms to fit the astrometric observations of this source, including a possible single companion (single companion solution). First we used both methods to fit the new VLBA astrometric observations to obtain proper motions and parallax, taking into account a single companion. 
Table~\ref{tab_4} summarizes the best fit and the $\chi^{2}_{red}$ per degree of freedom 
($\chi^{2}_{red}$ = $\chi^{2}/(N_{data} - N_{par} - 1)$, where $N_{data} = 2 \times N_{points}$ and $N_{par}$ is the number of fitted parameters). 
The fits of the parallax, proper motions, and orbital motions of the candidate are presented in Figure~\ref{fig_4}. 
The fit of the astrometric data clearly improves when including a companion, as seen by the $\chi^{2}_{red}$. The $\chi^{2}_{red}$ is now about a factor of 2 smaller, compared with the single-source solution.
Tables~\ref{tab_2} and \ref{tab_4} and Figures~\ref{fig_2} and \ref{fig_4} show that the residuals of the single-companion solution (RMS $\sim$ 0.10 mas) are a factor of 1.4 smaller than in the case of the single-source solution (RMS $\sim$ 0.14 mas). 
The residuals are now comparable to the mean noise present in the data (RMS $\sim$ 0.08 and 0.13 mas for both R.A. and decl.)
and the astrometric precision expected with the VLBA ($<$ 96 $\mu$$as$).
The astrometric signal in the source due to the companion is 0.17 $\pm$ 0.10 mas, i.e., significant at 1.7$\sigma$.
Although the astrometric signal is small with a relatively large error, the same signal appears when we analyze both datasets using three different algorithms (periodogram, least-squares, and AGA). This indicates that this astrometric signal is real. 

Table~\ref{tab_4} summarizes the best fit of the new VLBA astrometric data, including a companion. 
The orbit of the companion has an orbital period $P$ $\sim$ 241 days, a position angle of the line of nodes $\Omega$ $\sim$ 122$^\circ$, and an inclination angle $i$ $\sim$ 88$^\circ$, which indicates that the orbit of the companion is prograde ($i$ $<$ 90$^\circ$).
However, the large error in the inclination angle ($\sim$ 36$^\circ$) suggests that the orbit could also be retrograde ($i$ $>$ 90$^\circ$). In addition, the astrometric fit of the data indicates that the eccentricity of the orbit is not well constrained; thus, we use a fixed eccentricity, $e$ = 0. The orbital period obtained with the astrometric fit is consistent with that obtained with the periodogram. The orbit of the companion  is relatively well fitted; however, the errors of the orbital parameters are large. This is consistent with the relatively broad signal observed in the periodogram (see Figure~\ref{fig_3}). Further observations are needed to better constrain the orbital parameters of this companion. 
With this astrometric  fit, we cannot estimate the dynamical mass of the system; thus to estimate the mass of the companion, we use the lower and upper limits of the best estimated mass for this source $M_{*} = 0.06$$-$$0.08$ $M_\odot$  \citep[][]{martin94, reid02, hallinan08} as a fixed mass. Table~\ref{tab_4} summarizes the estimated parameters of the companion, hereafter TVLM~513$b$. We find that the mass of the companion is between 0.00036 ($M_{*} = 0.06$ $M_\odot$) and 0.00044 $M_\odot$ ($M_{*} = 0.08$ $M_\odot$), which is consistent with a planetary companion with a mass between 0.38 and 0.46 $M_{J}$. The semimajor axis of the orbit of this planetary companion is between 0.295 and 0.325 au.

We also fitted the combined VLBA data, including the Keplerian fit of a single companion (single-companion solution). Table~\ref{tab_4} summarizes the best fit and the $\chi^{2}_{red}$ per degree of freedom. 
The fit of the combined data also improves when including a companion. The residuals of the single-companion solution (RMS $\sim$ 0.13 mas) improve by a factor of 1.3 compared to the case of a single-source solution (RMS $\sim$ 0.17 mas). Similarly, the $\chi^{2}_{red}$ is smaller by a factor of 1.7 compared to the single-source solution.
The orbital fit to the combined data is in general similar to that obtained in the case of the fit of the new VLBA astrometric data (see Table~\ref{tab_4} and Figures~\ref{fig_4}). The orbital parameters and their estimated errors are similar, with relatively small differences from those obtained using only the new VLBA data (see Table~\ref{tab_4}). The estimated mass and the semimajor axis of the orbit of the companion are also similar to those obtained with the new VLBA data. These results further support the detection of a planetary companion.

These results indicate that the best fit of the orbit of the companion is obtained with the new VLBA data. This is not surprising because these observations are in general deeper and with smaller error bars than previous VLBA observations. However, it is important to point out that the astrometric signal appears in both the new VLBA data and the combined data. Furthermore, the same astrometric signal is found using the two different algorithms (least-squares and AGA) that we have used here, which is also consistent with the astrometric signal found in the least-squares periodogram.
Figures~\ref{fig_4} suggests a reasonably good astrometric fit. However, the residuals, although small (RMS $\sim$ 0.1 mas), are comparable to the astrometric signal (0.17 mas).  

Figures~\ref{fig_4} shows that the data are well fitted when considering a single planetary companion; however, Table~\ref{tab_4} shows that the orbital parameters obtained with the fit have large error bars, which indicates that the orbital motion of the companion is not well constrained. 
We find that this is the result of several contributions, which combined increase the astrometric errors and produce a larger error in the orbital fit. The astrometric signal of the companion is small (0.17 mas), just larger than both the residuals of the fit (RMS $\sim$ 0.1 mas) and the mean noise present in the data (RMS $\sim$ 0.08 and 0.13 mas for R.A. and decl.). In addition, the periodogram of the data (see Figures~\ref{fig_3}) indicates that the orbit of the companion is not completely constrained, and that there may be a second companion with a larger orbital period. The presence of a second companion would appear in the residuals as an additional source of noise, and it would worsen the astrometric fit of the detected companion. Thus, all of these contributions preclude a better estimate of the orbital parameters. However, the fact that the astrometric signal appears in the periodogram and in the fits obtained with two different algorithms (least-squares and AGA) supports the detection of the planetary companion.
Further observations are needed to better constrain the orbital solution and possibly to confirm the presence of the putative second companion.

\subsection{Distance to TVLM~513}

Table~\ref{tab_4}  shows that the estimated parallax to TVLM~513 does not change substantially when fitting only the new VLBA data or combining the new data with the previous VLBA data. Taking into account that the different fits give slightly different values for the parallax, we have calculated the weighted average of the estimated parallax as follows:

\begin{equation}
< \Pi  > = \frac{ \sum^{N}_{i} \pi_{i} / \sigma^{2}_{i} }  { \sum^{N}_{i} 1 / \sigma^{2}_{i} },
\end{equation}

\noindent
and the uncertainty is:

\begin{equation}
\sigma( < \Pi  >) = \sqrt{ \frac{ \sum^{N}_{i} (1/ \sigma^{2}_{i}) (\pi_{i} - <\Pi>)^2 }  { \sum^{N}_{i} 1 / \sigma^{2}_{i} } },
\end{equation}

\noindent
where $\pi_{i}$ and $\sigma_{i}$ are the estimated parallax of each fit and its uncertainty, respectively.

We obtain that the weighted parallax is of 93.368 $\pm$ 0.039 mas, which corresponds to a weighted distance $d = 10.7102\pm0.0045$ pc.
The estimated error corresponds to the standard deviation of the fitted values, which better reflects the dispersion seen in the different astrometric fits.
This estimate is an improvement on the distance to this source of 10.762 $\pm$ 0.027 pc, previously obtained with  VLBI observations \citep{forbrich13, gawronski17}. 
The estimated error that we obtain here is about a factor of 10 smaller than those obtained previously.  This is mainly due to the larger number of observations used for the present astrometric fit, the accuracy of the observations we present here (see Table~\ref{tab_1}) and a better coverage of the parallax.

\subsection{Proper Motions}

Table~\ref{tab_4}  also shows that the estimated proper motions of TVLM~513 do not change substantially when fitting the new VLBA data and the combined data. The fit of the combined VLBA data gives slightly better estimates for the proper motions because they cover a larger time span. We obtain that the weighted average proper motions are 
$\mu_{\alpha} = -43.164 \pm 0.011$ mas yr$^{-1}$ and 
$\mu_{\delta} = -65.528 \pm 0.010$ mas yr$^{-1}$.

\subsection{Expected RV}

The solution that we obtain for the single-companion astrometry can be used to estimate an expected induced RV of the star due to the gravitational pull of the companion as follows \citep[e.g.,][]{canto09, curiel19}:

\begin{equation}
K =\left(\frac{2 \pi G}{T}\right)^{1/3} \frac{m_{p} sin(i)} {(M_{*} + m_{p})^{2/3}} \frac{1} {\sqrt{1 - e^{2}}}
\end{equation}

\noindent
where $G$ is the gravitational constant, and $T$, $M_{*}$, $m_{p}$, and $e$ are the orbital period, the star and companion masses, and the eccentricity of the orbit of the companion. Using the solutions of the astrometric fit given in Table~\ref{tab_4}, we obtain that the maximum  RV of TVLM~513 induced by TVLM~513$b$ is $K\sim$ 64$-$81 m s$^{-1}$ (for the combined and the  new VLBA data, respectively). This RV could, in principle, be observed with high spectral resolution spectrographs.
Future short- and long-term high-resolution spectroscopic observations of TVLM~513 may be able to detect the RV signal that we find that TVLM~513$b$ induces on TVLM~513. Furthermore, these kinds of observations may also be able to confirm the putative second companion suggested by the periodogram.

\subsection{Flux variability of the source}

\begin{figure*}
   \centering
   \plotone{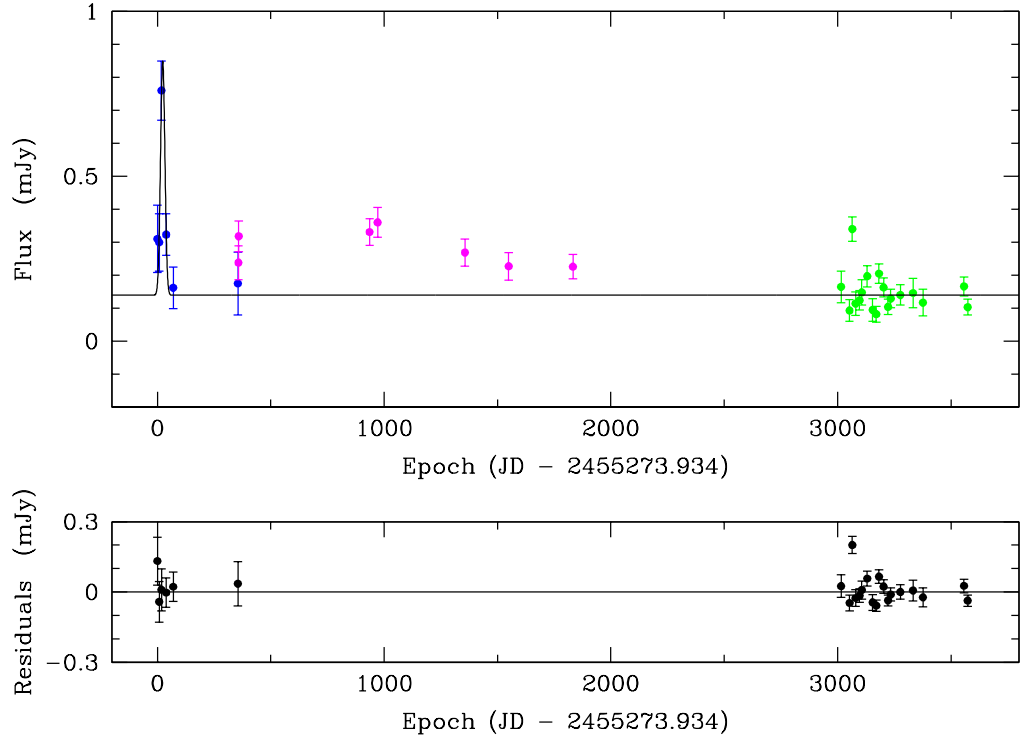}
    \caption{Radio flux density of TVLM~513 as a function of time. This figure includes all of the VLBI observations obtained with the VLBA (blue and green) and EVN (magenta). The VLBA observations were obtained at a frequency of 8.4 GHz, while the EVN observations were obtained at a frequency of 5 GHz. The flux density of the source presents short-term temporal flux density variations and seems to have a general tendency to decrease as function of time. The solid line corresponds to the fit of the data obtained with the VLBA. The lower panel shows the residuals of the fit, showing that the outburst observed about 10 yr ago is well fitted with a single gaussian function.}
    \label{fig_5}%
    \end{figure*}

The radio continuum flux density of TVLM~513 is clearly variable in time. Figure~\ref{fig_5} shows that the flux density of this source has short-term, and probably long-term, variability. The time scale of the short-term variability is of a few days, where the flux changes by about a factor of 2 or so. This flux variability is observed with both the VLBA observations obtained at 8.4 GHz  
and the European VLBI Network (EVN) observations obtained at 5 GHz obtained by \cite{gawronski17}.
This source is well known to be variable on a time scale of $\sim$ 1.96 hr (when polarized bursts of emission occur), which has been inferred to be the rotation period of this UCD   \citep[e.g.,][]{osten06, hallinan07, hallinan08, berger08}. These variability time scales may be unrelated, since the integration time of the VLBI observations is of several hours ($\sim$4.0 and 7.5 hr);  thus, a single VLBI observation integrates over several of the very short time span flux variations. 

In the long term, the flux density of the source has in general decreased as function of time in the past 10 yr. The source was in general stronger in the first VLBA-observed epochs and weaker in the last VLBA-observed epochs.  This suggests that the flux density of the source may have a general tendency to decrease as function of time, at least in the past 10 yr.
To investigate the long-term variability, and just for description purposes, we have fitted the data with a function of the type

\begin{equation}
Flux = f_{0} + f_{m}\ e^{-\frac{(t - t_{0})^{2}}{2 \sigma^{2}}}.
\end{equation}

\noindent
This function fits a single gaussian function plus a flux base to all of the VLBA epochs. Here $f_{0}$ is a constant flux density (in mJy), $f_{m}$ is the maximum increment in the flux density (in mJy) during the outburst, $t_{0}$ is the time of the maximum flux of the source during the outburst (in days from the first observed epoch), and $\sigma$ is the FWHM (in days) of the gaussian function.
In this fit, the single outburst observed about 10 yr ago is fitted with a single gaussian function.
The fit of the observed flux density data gives $f_{0}$ = 0.14 mJy, $f_{m}$ = 0.71 mJy, $t_{0}$ = 23.21 days, and $\sigma$ = 9.60 days. 
This fit suggests that the source had a maximum flux outburst of about 0.85 mJy centered at the epoch JD = 2,455,297.139, which is very close to the third observed epoch (BF100C; see Table~\ref{tab_1}), and that the outburst had an FWHM of 9.60 days and lasted for about 70 days.
For the fit, we have only used the observations obtained with the VLBA because they were obtained at 8.4 GHz, while the EVN observations were obtained at 5 GHz.
In Figure~\ref{fig_5}  we plot the observed flux densities, the fit that we obtain, and the residuals of the fit. 
The residuals of the fit show that the previous VLBA observations can be well fitted with a single gaussian function with an amplitude of about 0.71 mJy and an FWHM of about 9.60 days, and that the new VLBA observations do not show a similar outburst.
The fit also shows that the source is generally weak, having a mean flux density around 0.14 mJy with a small flux fluctuation of about 0.1 mJy in short periods of time, probably of a few days, or even shorter. The source may also have strong outbursts, such as the one observed about 10 yr ago that lasted for about 70 days (see Figure~\ref{fig_5}). The large temporal gap of about 7 yr in the VLBA data precludes the possibility of finding whether these outbursts may be periodic or not. Our recent VLBA observations, which were obtained in a time span of about 560 days, do not show any strong outbursts, suggesting that if the source undergoes periodic outbursts, they are probably at intervals longer than this time scale. Thus, we find that the source seems to undergo flux fluctuations with at least three different time spans: a) a short-period variation with a time span of about 1.96 hr, observed with the VLA and correlating with the rotation period of this UCD; b) an intermediate-period variation with a time span of a few days (not well established), observed with the VLBA; and c) a possible longer-period variation, observed with the VLBA as a single outburst about 10 yr ago. Future observations will tell whether the source undergoes periodic outbursts, as that observed about 10 yr ago, and what their origin may be.

\subsection{First exoplanet found with radio astrometry}

There is only one exoplanet that has been found using  astrometry  \citep[HD 176051~b;][]{muterspaugh10}. It was found using optical differential astrometry. This planetary companion is associated with a relatively nearby (14.99 $\pm$ 0.13 pc) binary system (1.07 and 0.71 $M_\odot$), and has an estimated mass of 1.5 $\pm$ 0.3 $M_{J}$, assuming that it is associated with the low mass star. The mass of the planetary companion is expected to be higher if it is associated to the higher-mass star.

\begin{figure*}
   \centering
   \plottwo{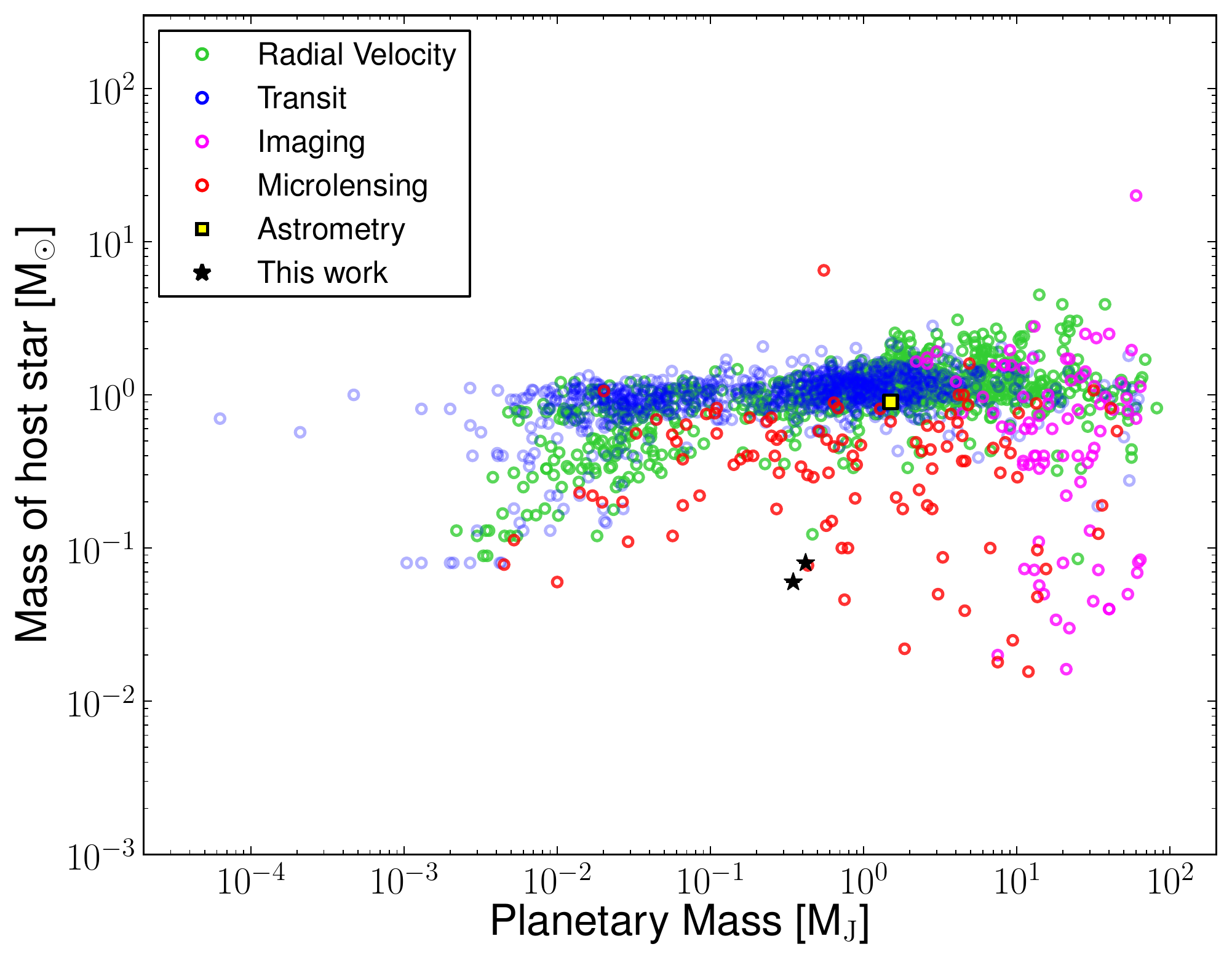}{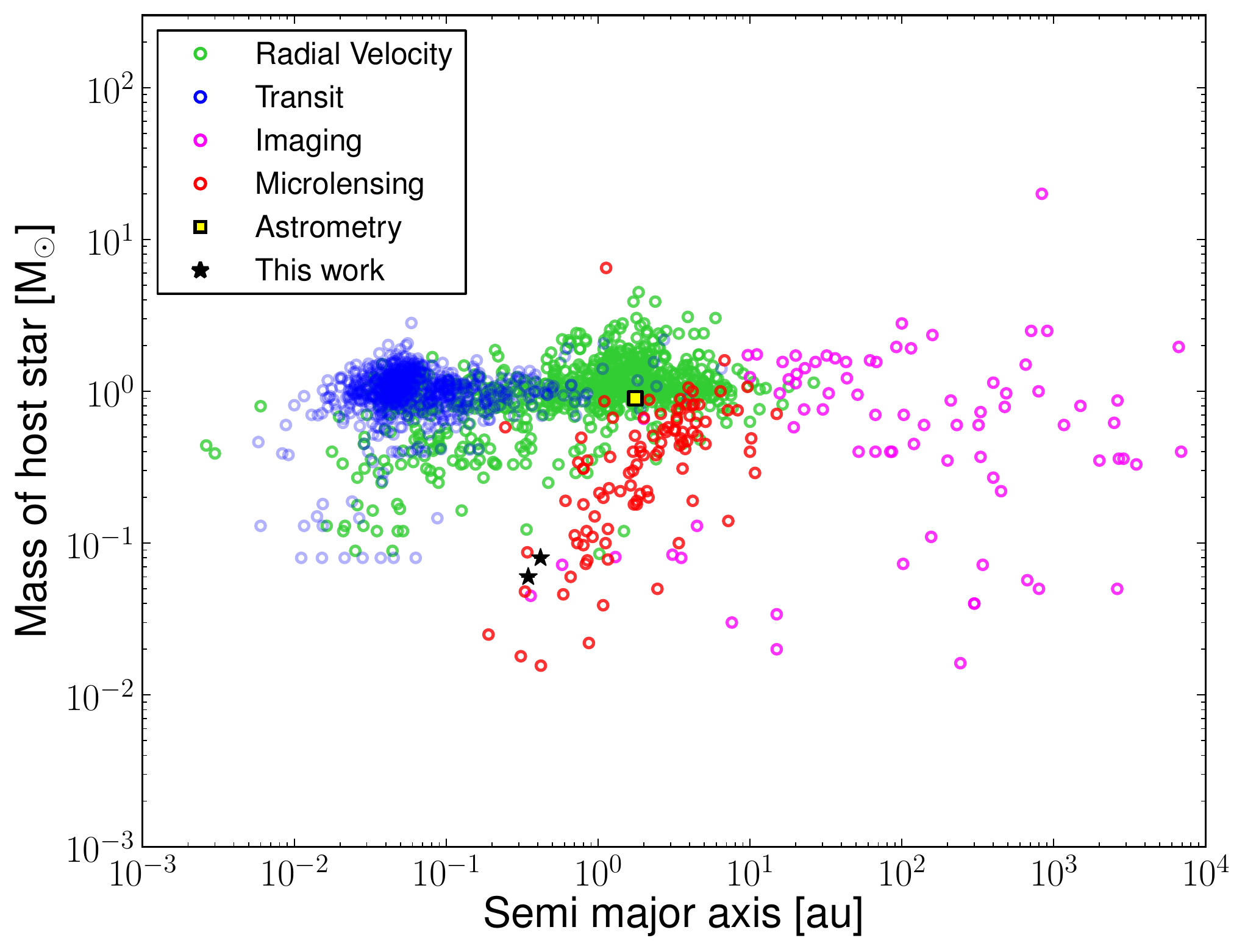}
    \caption{Distribution of the planetary mass and the semimajor axis of the planetary orbit $vs.$ the mass of the host star of known exoplanets and candidates as listed at exoplanet.eu  \citep{schneider11}. The five main detection methods are marked by different colors. The two black stars indicate the position of TVLM~513$b$ for the estimated mass limits of the M9~UCDTVLM~513.} 
    \label{fig_6}%
    \end{figure*}

The best fit of the astrometric data of the M9~dwarf TVLM~513 indicates that this UCD has at least one substellar companion, TVLM~513$b$. Furthermore, the estimated weighted average mass and semi-major axis of TVLM~513$b$ are 0.347 $\pm$ 0.035 $M_{J}$, and 0.2789 $\pm$ 0.0034 au, respectively, when assuming the lower mass limit of TVLM~513 (0.06 $M_\odot$), or 0.418 $\pm$ 0.040 $M_{J}$, and 0.3072 $\pm$ 0.0040 au, respectively, when assuming the upper mass limit of TVLM~513 (0.08 $M_\odot$). 
The estimated weighted average period and inclination angle of the orbit are 221 $\pm$ 5 days and 80 $\pm$ 9$^\circ$, respectively.
The estimated mass is consistent with this planetary companion being a Saturn-mass planet (0.30 $M_{J}$).
Figure~\ref{fig_6} shows all the confirmed planets that have been found up to now for which the planetary mass has been estimated (either $m_{p}$ or $m_{p} \times sin(i)$). We include TVLM~513$b$ in this figure. 
This figure shows that TVLM~513$b$ is located in a region in the $M_{*} - m_{p}$ and $M_{*} - a_{p}$ phase space where very few planets have been found. TVLM~513 is one of the lowest-mass stars with known Jovian-mass planetary companions.

The estimated weighted average astrometric signal of TVLM~513 is 0.145 $\pm$ 0.019 mas. Although this astrometric signal is relatively small, it is consistent with a planetary companion associated with this M9 UCD. However, this astrometric signal could be contaminated by the expected astrophysical $''$jitter$''$ added to the true source position due to stellar activity. It is estimated that M9 UCD have stellar radius of $\sim$ 0.1 R$_\sun$ \citep[e.g.,][]{chabrier00b,dahn02,hallinan06}.
Thus, assuming that the radio emission is originated within $\sim$1 stellar radius of the photosphere  \citep[e.g.,][]{white94}, the expected radius of TVLM~513 at a distance of 10.7 pc is about 0.05 mas. Thus, the expected jitter is about a factor of 3 smaller than the astrometric signal observed in TVLM~513. This result supports the detection of the planetary companion TVLM~513$b$.
 
As we have mentioned before, in recent years, it has been found that giant-mass planets, such as the one we have found orbiting TVLM~513, have a very low occurrence around UCDs, which is consistent with predictions of planetary formation theories. The core-accretion theory predicts that the formation of giant-mass planets scales with the mass of the central star; thus, it is expected that very few Jovian-mass planets are formed around UCDs \citep[e.g.,][]{laughlin04, kennedy08}. The core-accretion theory indicates that these planets would be formed in orbits far from the star, at several au. On the other hand, it is expected that disk instability may also be able to form giant-mass planets around UCDs  \citep[e.g.,][]{boss06}. In this case, the orbit of the planet is expected to be relatively closer to the star, from a few to several au. The semimajor axis of the orbit of TVLM~513$b$, $a \sim$ 0.3 au, is smaller than expected from core-accretion and disk instability theoretical models \citep[e.g.,][]{laughlin04, boss06}. It may be that TVLM~513$b$ was formed by the same collisional accumulation process that led to the formation of the terrestrial planets in our solar system. Alternatively, TVLM~513$b$ may have formed with a wider orbit, at several au from the star, and then migrated inward to its current orbit. However, it is not clear what would stop the migration of the planet at 0.3 au. Further theoretical models will be required to understand the formation of giant-mass planets, such as the one we find associated with the M9 UCD TVLM~513.

Finally, to our knowledge, this is the second exoplanet found using astrometry and the first exoplanet found using absolute astrometry. In addition, this is also the first exoplanet found using radio astrometric observations. This result suggests that radio observations with the VLBA can be used to search for giant-mass planets around very low mass stars, such as M dwarfs, and in particular around UCDs.

\section{Conclusions}

The multi-epoch VLBA observations of the M9~dwarf TVLM~513 that we present here allow us to carry out a precise analysis of the spatial wobbling of this source due to its parallax and its proper motions, as well as to search for possible companions. The precise astrometric observations obtained with the VLBA were crucial to carry out this kind of study. We find that the determination of the distance to this source improves significantly.

Here we present different ways to analyze the VLBA astrometric observations of the M9~dwarf TVLM~513. We have used two different algorithms (a least-squares algorithm and a genetic algorithm) to fit the astrometric multi-epoch data obtained with the VLBA. First, we only fit the parallax  and proper motions of the host star. The residuals of this fit are large compared with the noise of the observed data and the expected precision of the multi-epoch VLBA observations. 

We have searched for possible companions using a recursive least-squares periodogram, finding a companion candidate in the periodogram. 
We also find that the astrometric fit improves substantially when including the orbit of a companion in the fit. We find that the parameters of the orbit are consistent with a planetary companion of 0.347 $\pm$ 0.035 $M_{J}$, with an orbital period of 221 $\pm$ 5 days and having a semimajor axis of  0.2789 $\pm$ 0.0034 au, assuming the estimated lower mass limit of TVLM~513 (0.06 $M_\odot$), or 0.418 $\pm$ 0.040 $M_{J}$, with an orbital period of 221 $\pm$ 5 days and a semimajor axis of 0.3072 $\pm$ 0.0040 au, assuming the estimated upper mass limit of TVLM~513 (0.08 $M_\odot$). The estimated orbital motions of TVLM~513$b$ are consistent with being a Saturn-like planet in a compact, probably circular orbit and with a large inclination angle ($\sim$ 80$^\circ$). This is the second exoplanet found with the astrometry technique and the first exoplanet found using absolute astrometry. It is also the first exoplanet that has been found with radio astrometric observations.

\begin{acknowledgements}
We thank the reviewer for his/her valuable comments that helped to improve this paper. 
S.C. acknowledges support from UNAM and CONACyT, M\'exico. 
This work was supported by  UNAM-PAPIIT IN103318.
G.N.O.-L. acknowledges support from the von Humboldt Stiftung.
The Long Baseline Observatory is a facility of the National Science Foundation operated under a cooperative agreement by Associated Universities, Inc. The National Radio Astronomy Observatory is a facility of the National Science Foundation operated under a cooperative agreement by Associated Universities, Inc.
\end{acknowledgements}

~~
~~
~~

{~~~~~~~~~ORCID iDs}

Salvador Curiel https:/orcid.org/0000-0003-4576-0436

Gisela N. Ortiz-Le\'on https:/orcid.org/0000-0002-2863-676X

Rosa M. Torres https:/orcid.org/0000-0002-7179-6427


\bibliographystyle{aasjournal}

\begin{thebibliography}{}

\bibitem[Adam et al.(2005)]{adam05} Adams, F. C., Bodenheimer, P., \& Laughlin, G. 2005,  Astron. Nachr., 326, 913 
\bibitem[Andrei et al.(2011)]{andrei11} Andrei, A. H., Smart, R. L., Penna, J. L., et al. 2011,  \aj, 141, 54
\bibitem[Anglada-Escud\'e \& Tuomi.(2012)]{anglada12} Anglada-Escud\'e, G., \& Tuomi, M.  2012,  \aap, 548, A58
\bibitem[Berger(2002)]{berger02} Berger, E. 2002,  \apj, 572, 503 
\bibitem[Berger et al.(2008)]{berger08} Berger, E., Gizis, J. E., Giampapa, M. S., et al.  2008,  \apj, 673, 1080 
\bibitem[Blake et al.(2010)]{blake10} Blake, C. H., Charbonneau, D., \& White, R. J. 2010,  \apj, 723, 684    
\bibitem[Bonfils et al.(2013)]{bonfils13} Bonfils, X., Delfosse, X., Udry, S., et al.  2013,  \aap, 549, A109 
\bibitem[Boss(2006)]{boss06} Boss, A. P. 2006,  \apj, 643, 501 
\bibitem[Boss et al.(2009)]{boss09} Boss, A. P., Weinberger, A. J., Anglada-Escud\'e, G., et al. 2009,  \pasp, 121, 1218 
\bibitem[Bower et al.(2009)]{bower09} Bower, G. C., Bolatto, A., Ford, E. B., \& Kalas, P. 2009,  \apj, 701, 1922 
\bibitem[Bower et al.(2011)]{bower11} Bower, G. C., Bolatto, A., Ford, E. B., et al. 2011,  \apj, 740, 32
\bibitem[Cant\'o et al.(2009)]{canto09} Cant\'o, J., Curiel, S, \& Mart{\'\i}nez-G\'omez, E. 2009, \aap, 501, 1259 
\bibitem[Casertano et al.(2008)]{casertano08} Casertano, S., Lattanzi, M. G., Sozzetti, A., et al. 2008, \aap, 482, 699
\bibitem[Chabrier \& Baraffe(2000)]{chabrier00} Chabrier, G. \& Baraffe, I.  2000,  \araa, 38, 337 
\bibitem[Chabrier etal.(2000)]{chabrier00b} Chabrier, G., Baraffe, I., Allard, F. \& Hauschildt, P.  2000,  \apj, 542, 464 
\bibitem[Close et al.(2003)]{close03} Close, L. M., Siegler, N., Freed, M., \& Biller, B. 2003, \apj, 587, 407  
\bibitem[Cumming(2004)]{cumming04} Cumming, A. 2004, \mnras, 354, 1165  
\bibitem[Cumming et al.(2008)]{cumming08} Cumming, A., Butler, R. P., Marcy, G. W., et al. 2008,  \pasp, 120, 531
\bibitem[Curiel et al.(2019)]{curiel19} Curiel, S., Ortiz-Le\'on, G. N., Mioduszewski, A.J., \& Torres, R. M. 2019,  \apj, 884, 13 
\bibitem[Dahn et al.(2002)]{dahn02} Dahn, C. C., Harris, H. C., Vrba, F. J., et al. 2002,  \aj, 124, 1170 
\bibitem[Dressing \& Charbonneau(2013)]{dressing13} Dressing, C. D. \& Charbonneau, D. 2013,  \apj, 767, 95     
\bibitem[Dupuy \& Kraus(2013)]{dupuy13} Dupuy, T. J., \& Kraus, A. L. 2013,  Sci, 341, 1492 
\bibitem[Dupuy \& Liu(2012)]{dupuy12} Dupuy, T. J., \& Liu, M. C. 2012,  \apjs, 201, 19 
\bibitem[Dupuy et al.(2016)]{dupuy16} Dupuy, T. J., Forbrich, J, Rizzuto, A. et al. 2016,  \apj, 827, 23 
\bibitem[Endl et al.(2006)]{endl06} Endl, M., Cochran, W. D., K\"urster, M., Paulson, D. B., et al. 2006,  \apj, 649, 436   
\bibitem[Forbrich et al.(2013)]{forbrich13} Forbrich, J., Berger, E., \& Reid, M. J. 2013, \apj, 777, 70
\bibitem[Forbrich et al.(2016)]{forbrich16} Forbrich, J., Dupuy, T. J., Reid, M. J., et al. 2016, \apj, 827, 22
\bibitem[Gawro{\'n}ski et al.(2017)]{gawronski17} Gawro{\'n}ski, M. P., Go{\'z}dziewski, K., \& Katarzy{\'n}ski, K. 2017, \mnras, 466, 4211
\bibitem[Gillon et al.(2017)]{gillon17} Gillon, M., Triaud, A. H. M. J., Demory, B. O. et al. 2017,  Natur, 542, 456  
\bibitem[Greisen(2003)]{greisen03} Greisen, 2003, in Information Handling in Astronomy: Historical Vistas, Vol. 285, ed. A. Heck (New York: Springer), 109
\bibitem[Hallinan et al.(2006)]{hallinan06} Hallinan, G., Antonova, A., Doyle, J. G.. et al. 2006,  \apj, 653, 690 
\bibitem[Hallinan et al.(2007)]{hallinan07} Hallinan, G., Brouke, S., Lane, C. et al. 2007,  \apj, 663, L25 
\bibitem[Hallinan et al.(2008)]{hallinan08} Hallinan, G., Antonova, A., Doyle, J. G.. et al. 2008,  \apj, 684, 644 
\bibitem[Ida \& Lin(2005)]{ida05} Ida, S., \& Lin, D. N. C. 2005,  \apj, 626, 1045 
\bibitem[Kennedy \& Kenyon(2008)]{kennedy08} Kennedy, G. M., \& Kenyon, S. J. 2008,  \apj, 673, 502 
\bibitem[Kubas et al.(2012)]{kubas12} Kubas, D., Beaulieu, J. P., Bennett, D. P. 2012,  \aap, 540, A78 
\bibitem[Laughlin et al.(2004)]{laughlin04} Laughlin, G., Bodenheimer, P., \& Adams, F. C. 2004,  \apjl, 612, L73   
\bibitem[Leggett(1992)]{leggett92} Leggett, S. K. 1992,  \apjs, 82, 351   
\bibitem[Leggett et al.(1998)]{leggett98} Leggett, S. K., Allard, F., \& Hauschildt, P. H. 1998,  \apj, 509, 836
\bibitem[Mart{\'\i}n et al.(1994)]{martin94} Mart{\'\i}n, E. L., Rebolo, R. \& Magazz\'u 1994,  \apj, 436, 262 
\bibitem[Muirhead et al.(2012)]{muirhead12} Muirhead, P. S., Johnson, J. A., Apps, K., et al. 2012,  \apj, 747, 144 
\bibitem[Muterspaugh et al.(2010)]{muterspaugh10} Muterspaugh, M. W., Lane, B. F., Kulkarni, S. R., et al. 2010,  \aj, 140, 1657 
\bibitem[Ortiz-Le\'on et al.(2017)]{ortizleon17} Ortiz-Le\'on, G. N., Loinard, L., Kounkel, M. A., et al. 2017, \apj, 834, 1410
\bibitem[Osten et al.(2006)]{osten06} Osten, R. A., Hawley, S. L., Bastian, T. S., \& Reid, I. N. 2006,  \apj, 637, 518 
\bibitem[Perryman et al.(2014)]{perryman14} Perryman, M., Hartman, J., Bakos, G. \'A., \&  Lindegren, L. 2014, \apj, 797, 14
\bibitem[Pravdo \& Shaklan(1996)]{pravdo96} Pravdo, S. H., \& Shaklan, S. B. 1996,  \apj, 465, 264 
\bibitem[Reid et al.(2008)]{reid08} Reid, I. N., Cruz, K. L., Kirkpatrick, J. D., et al. 2008,  \aj, 136, 1290
\bibitem[Reid et al.(2002)]{reid02} Reid, I. N., Kirkpatrick, J. D., Liebert, J., et al. 2002,  \aj, 124, 519  
\bibitem[Rodler et al.(2012)]{rodler12} Rodler, F., Deshpande, R., Zapatero-Osorio, M.R., et al. 2012,  \aap, 538, A141  
\bibitem[Sahlmann(2012)]{sahlmann12} Sahlmann, J. 2012, Ph.D. Thesis, Observatoire de Gen\`eve, Universit\'e de Gen\`eve 
\bibitem[Sahlmann et al.(2013)]{sahlmann13} Sahlmann, J., Lazorenko, P. F., S\'egransan, D., et al.  2013, \aap, 556, A133
\bibitem[Sahlmann et al.(2014)]{sahlmann14} Sahlmann, J., Lazorenko, P. F., S\'egransan, D. et al. 2014,  \aap, 565, A20   
\bibitem[Sahlmann et al.(2016)]{sahlmann16} Sahlmann, J., Lazorenko, P. F., S\'egransan, D., et al. 2016, \aap, 595, A77
\bibitem[Scargle(1982)]{scargle82} Scargle, J. D. 1982, ApJ, 263, 835   
\bibitem[Schneider et al.(2011)]{schneider11} Schneider, J., Dedieu, C., Le Sidaner, P., Savalle, R., \& Zolotukhin, I.. 2011, \aap, 532, A79
\bibitem[Smart et al.(2013)]{smart13} Smart, R. L., Tinney, C. G., Bucciarelli, B., et al. 2013,  \mnras, 433, 2054   
\bibitem[Sozzetti(2005)]{sozzetti05} Sozzetti, A. 2005,  \pasp, 117, 1021 
\bibitem[Sozzetti et al.(2014)]{sozzetti14} Sozzetti, A., Giacobbe, P., Lattanzi, M. G., et al. 2014, \mnras, 437, 497.
\bibitem[Stumpf et al.(2010)]{stumpf10} Stumpf, M. B., Brandner, W., Joergens, V. et al.  2010,  \apj, 724, 1-11
\bibitem[Thompson et al.(2017)]{thompson17} Thompson, A. Richard,  Moran, James M., \&
Swenson, George W., Jr. 2017, Interferometry and Synthesis in Radio Astronomy, 3rd Edition
\bibitem[West et al.(2011)]{west11} West, A. A., Morgan, D. P., Bochanski, J. J., et al. 2011, \aj, 141, 97   
\bibitem[White et al.(1994)]{white94} White, S. M., Lim, J. \& Kundu, M. R. 1994,  \apj, 422, 293   


\end{thebibliography}


{}


\clearpage



\begin{deluxetable*}{ccclccccccc}
\tabletypesize{\scriptsize}
\tablewidth{0pt}
\tablecolumns{11}
\tablecaption{Observed Epochs and Measured VLBA Positions \label{tab_1} }
\tablehead{ \\ Project & Date & Start UT &     Stop UT  & Julian Date &      $\alpha$ (J2000)     &  $\sigma_\alpha$   & $\delta$ (J2000)  & $\sigma_\delta$  & rms ($\mu$Jy) & Flux Density ($\mu$Jy) \\
                   (1)        &        (2)       &        (3)                           &            (4)              &                (5)        &  (6)     &   (7)  & (8) & (9) & (10) & (11)  \\
}
\startdata
BF100A	&	2010 Mar 18	&	07:57:38	&	12:51:46	&	2,455,273.9338	&	15	01	8.15964647	&	0.00000853 	&	22	50	1.4994274	&	0.0001280	&	49	&	310	$\pm$	102 \\
BF100B	&	2010 Mar 26	&	07:26:10	&	12:28:47	&	2,455,281.9149	&	15	01	8.15896008	&	0.00000688 	&	22	50	1.5057512	&	0.0001032	&	50	&	300	$\pm$	87 \\
BF100C	&	2010 Apr 05	&	06:46:53	&	11:49:30	&	2,455,291.8876	&	15	01	8.15800042	&	0.00000363 	&	22	50	1.5132772	&	0.0000545	&	52	&	760	$\pm$	90 \\
BF100I	&	2010 Apr 26	&	04:59:27	&	11:59:18	&	2,455,312.8537	&	15	01	8.15568955	&	0.00000601 	&	22	50	1.5242912	&	0.0000901	&	31	&	323	$\pm$	63 \\
BF100D	&	2010 May 27	&	03:22:24	&	08:25:03	&	2,455,343.7456	&	15	01	8.15206035	&	0.00000686 	&	22	50	1.5282724	&	0.0001029	&	52	&	162	$\pm$	63 \\
BF100E	&	2010 Jun 25	&	01:28:24	&	06:31:01	&	2,455,372.6665	&		--		&	--	&		--		&	--	&	54	&	-- \\
BF100F	&	2010 Nov 03	&	16:49:22	&	21:52:01	&	2,455,504.3060	&		--		&	--	&		--		&	--	&	52	&	-- \\
BF100G	&	2011 Mar 08	&	08:39:26	&	13:43:03	&	2,455,628.9661	&	15	01	8.15725541	&	0.00001483 	&	22	50	1.4251275	&	0.0002224	&	59	&	175	$\pm$	95 \\
BF100H	&	2011 Aug 03	&	22:53:34	&	03:57:13(+1)	&	2,455,777.5593	&		--		&	--	&		--		&	--	&	54	&	-- \\
BC236A	&	2018 Jun 20	&	00:44:26	&	07:25:18	&	2,458,289.6701	&	15	01	8.12460313	&	0.00000913 	&	22	50	0.9942284	&	0.0001370	&	25	&	165	$\pm$	48 \\
BC236B	&	2018 Jul 26	&	22:19:22	&	05:00:14(+1)	&	2,458,326.5693	&	15	01	8.12218516	&	0.00000849 	&	22	50	0.9628254	&	0.0001273	&	13	&	93	$\pm$	33 \\
BC244A	&	2018 Aug 07	&	21:32:38	&	04:13:30(+1)	&	2,458,338.5369	&	15	01	8.12191526 	&	0.00000290 	&	22	50	0.9492577	&	0.0000434	&	18	&	340	$\pm$	37 \\
BC236C	&	2018 Aug 22	&	20:33:42	&	03:14:33(+1)	&	2,458,353.4959	&	15	01	8.12193756	&	0.00000715 	&	22	50	0.9308916	&	0.0001073	&	18	&	114	$\pm$	36 \\
BC244B	&	2018 Sep 08	&	19:27:22	&	02:08:14(+1)	&	2,458,370.4499	&	15	01	8.12244619	&	0.00000504 	&	22	50	0.9098083	&	0.0000756	&	19	&	124	$\pm$	29 \\
BC236D	&	2018 Sep 18	&	18:48:02	&	01:28:55(+1)	&	2,458,380.4226	&	15	01	8.12298116	&	0.00000756 	&	22	50	0.8976996	&	0.0001134 	&	19	&	148	$\pm$	39 \\
BC244C	&	2018 Oct 12	&	17:13:49	&	23:54:41	&	2,458,404.3571	&	15	01	8.12482797	&	0.00000375 	&	22	50	0.8714270	&	0.0000563	&	19	&	197	$\pm$	32 \\
BC236E	&	2018 Nov 05	&	15:37:28	&	22:18:19	&	2,458,428.2902	&	15	01	8.12721613	&	0.00000787 	&	22	50	0.8530176	&	0.0001181 	&	18	&	95	$\pm$	34 \\
BC244D	&	2018 Nov 21	&	14:34:34	&	21:15:26	&	2,458,444.2465	&	15	01	8.12884899	&	0.00000644 	&	22	50	0.8454972	&	0.0000966	&	16	&	82	$\pm$	24 \\
BC236F	&	2018 Dec 03	&	13:47:57	&	20:28:48	&	2,458,456.2142	&	15	01	8.13003309	&	0.00000347 	&	22	50	0.8430444	&	0.0000520	&	15	&	205	$\pm$	29 \\
BC244E	&	2018 Dec 24	&	12:25:23	&	19:06:14	&	2,458,477.1568	&	15	01	8.13180704	&	0.00000433 	&	22	50	0.8456374	&	0.0000650	&	17	&	163	$\pm$	29 \\
BC236G	&	2019 Jan 12	&	11:10:41	&	17:51:33	&	2,458,496.1049	&	15	01	8.13288887	&	0.00000522 	&	22	50	0.8543747	&	0.0000782	&	12	&	104	$\pm$	23 \\
BC244F	&	2019 Jan 24	&	10:23:30	&	17:04:22	&	2,458,508.0722	&	15	01	8.13325331	&	0.00000486 	&	22	50	0.8626375	&	0.0000730	&	17	&	129	$\pm$	28 \\
BC236H	&	2019 Mar 08	&	07:34:26	&	14:15:18	&	2,458,550.9548	&	15	01	8.13228304	&	0.00000500 	&	22	50	0.9001560	&	0.0000751	&	17	&	140	$\pm$	31 \\
BC236I	&	2019 May 03	&	03:54:15	&	10:35:07	&	2,458,606.8019	&	15	01	8.12682413	&	0.00001012 	&	22	50	0.9363863	&	0.0001518	&	22	&	146	$\pm$	45 \\
BC236J	&	2019 Jun 16	&	01:02:20	&	07:43:12	&	2,458,650.6825	&	15	01	8.12189535	&	0.00001064 	&	22	50	0.9313250	&	0.0001596	&	20	&	117	$\pm$	40 \\
BC255A   &       2019 Dec 13     &      12:44:49   &       20:15:05  &       2,458,831.1875  &     15	01     8.12780012      &       0.00000465       &       22	50	0.7779006	&	0.0000697	& 	14	&	166	$\pm$	29 \\  
BC255B   &       2019 Dec 30     &      11:37:59   &       19:07:37  &       2,458,848.1408  &     15	01     8.12907848      &       0.00000607       &       22	50	0.7826089	&	0.0000911	& 	13	&	103	$\pm$	24 \\
\enddata
\end{deluxetable*}

%
\begin{deluxetable*}{lcc}
\centering                          
\tabletypesize{\scriptsize}
\tablewidth{0pt}
\tablecolumns{4}
\tablecaption{Single-source Astrometry Fits\tablenotemark{a} \label{tab_2}} 
\tablehead{
\\ Parameter & VLBA\_new  &  VLBA\_combined   
}
\startdata
&  Parameters Fitted   & \\
\hline                        
Epochs &  17  &   23     \\
$\mu_{\alpha}$ (mas yr$^{-1}$)  &   $-$43.21 $\pm$ 0.12   & $-$43.158 $\pm$ 0.012   \\
$\mu_{\delta}$ (mas yr$^{-1}$)  &  $-$65.37 $\pm$ 0.13    & $-$65.532 $\pm$ 0.013   \\
$\Pi$ (mas)                        &  93.450 $\pm$ 0.057     & 93.423 $\pm$  0.053   \\
\hline                        
&  Other Parameters & \\
\hline                        
$D$ (pc)                   &  10.7009 $\pm$ 0.0065      & 10.7040 $\pm$ 0.0061    \\
$\Delta$$\alpha$ (mas)\tablenotemark{b} & 0.079 & 0.099  \\
$\Delta$$\delta$ (mas)\tablenotemark{b}  & 0.123 & 0.138  \\
$\chi^2$, $\chi^2_{red}$   &   38.14, 1.36 &  66.18, 1.65   \\
%
\enddata
\tablenotetext{a}{The parameters presented here were obtained with AGA. Very similar results were obtained with the least-squares fitting method. The second column contains the astrometric fit of the new VLBA data. The third column corresponds to the astrometric fit of the combined VLBA data.}
\tablenotetext{b}{The rms dispersion of the residual.}
\end{deluxetable*}

%
\begin{deluxetable*}{lcc}
\centering                          
\tabletypesize{\scriptsize}
\tablewidth{0pt}
\tablecolumns{3}
\tablecaption{Single-source Astrometry Fits\tablenotemark{a} \label{tab_3}} 
\tablehead{                
\\ Parameter & VLBA\_new  &  VLBA\_combined   
}
\startdata
&  Parameters Fitted  & \\
\hline                        
Epochs &  17  &   23     \\
$\mu_{\alpha}$ (mas yr$^{-1}$)  &   $-$43.08 $\pm$ 0.12       & $-$43.187 $\pm$ 0.012   \\
$\mu_{\delta}$ (mas yr$^{-1}$)   &  $-$65.49 $\pm$ 0.13        & $-$65.460 $\pm$ 0.013  \\
$a_{\alpha}$ (mas yr$^{-2}$)      &  $-$0.34 $\pm$ 0.30          & $-$0.0144 $\pm$ 0.0045   \\
$a_{\delta}$ (mas yr$^{-2}$)       &   0.41 $\pm$ 0.33              & 0.0332 $\pm$ 0.0049          \\
$\Pi$ (mas)                                 &  93.407 $\pm$ 0.057          & 93.451 $\pm$  0.053   \\
\hline                        
&   Other Parameters  & \\
\hline                        
$D$ (pc)                                                     &  10.7058 $\pm$ 0.0065   & 10.7008 $\pm$ 0.0061    \\
$\Delta$$\alpha$ (mas)\tablenotemark{b} & 0.081 & 0.089  \\
$\Delta$$\delta$ (mas)\tablenotemark{b}  & 0.111 & 0.130  \\
$\chi^2$, $\chi^2_{red}$                             &   33.57, 1.29 &  53.87, 1.42    \\
\enddata
\tablenotetext{a}{The parameters presented here were obtained with AGA. Very similar results were obtained with the least-squares fitting method. The astrometric fit includes acceleration terms. The second column contains the astrometric fit of the new VLBA data. The third column corresponds to the astrometric fit of the combined VLBA data.}
\tablenotetext{b}{The rms dispersion of the residual.}
\end{deluxetable*}

%
%
\begin{deluxetable*}{lcc}
\centering
\tabletypesize{\scriptsize}
\tablewidth{0pt}
\tablecolumns{3}
\tablecaption{Single-companion Astrometry Fits\tablenotemark{a} \label{tab_4}}             
\tablehead{
\\ Parameter &   VLBA\_new & VLBA\_combined  
}
\startdata
  &    Parameters Fitted     &  \\
\hline                        
Epochs &  17  &   23    \\
%
$\mu_{\alpha}$ (mas yr$^{-1}$) &    $-$43.05 $\pm$ 0.17  & $-$43.165 $\pm$ 0.017   \\
$\mu_{\delta}$ (mas yr$^{-1}$)  &   $-$65.43 $\pm$ 0.18   & $-$65.529 $\pm$ 0.019   \\
$\Pi$ (mas)                                &   93.326 $\pm$ 0.079   & 93.405 $\pm$ 0.074          \\
$P$ (days)                                  & 241 $\pm$ 20     & 220 $\pm$ 5    \\
$T_0$ (days)                              & 2,458,263 $\pm$ 21   & 2,457,631 $\pm$  22   \\
$e$\tablenotemark{b}                                             & 0.0   &  0.0 \\ 
$\omega$ (deg)\tablenotemark{b}                         & 0.0       & 0.0      \\
$\Omega$ (deg)                        & 122 $\pm$ 35     & 139 $\pm$ 39   \\
$a_1$ (mas)                              & 0.17 $\pm$ 0.10           & 0.128 $\pm$ 0.088      \\
$i$ (deg)                                    & 88 $\pm$ 36       & 71 $\pm$ 38           \\
\hline                        
& Other Parameters   & \\
\hline                        
$D$ (pc)                   &   10.7151 $\pm$ 0.0091 & 10.7060 $\pm$ 0.0085   \\
$m  ~(M_\odot)$\tablenotemark{c}  & 0.08, 0.06    & 0.08, 0.06        \\
$m_2 ~(M_\odot)$ & 0.00044 $\pm$ 0.00023, 0.00036 $\pm$ 0.00019 &  0.00036 $\pm$ 0.00024, 0.00030 $\pm$ 0.00020 \\
$m_2 ~(M_{J})$    & 0.46 $\pm$ 0.25, 0.38 $\pm$ 0.20                          & 0.38 $\pm$ 0.24, 0.31 $\pm$ 0.21    \\
$a_1 ~(au)$         & 0.0018 $\pm$ 0.0011, 0.0018 $\pm$ 0.0011           & 0.00138 $\pm$ 0.00094, 0.00138 $\pm$ 0.00094  \\
$a_2 ~(au)$         & 0.325 $\pm$ 0.016, 0.295 $\pm$ 0.015                  & 0.3063 $\pm$ 0.0036, 0.2782 $\pm$ 0.0032  \\
$\Delta$$\alpha$ (mas)\tablenotemark{d} & 0.063 &  0.070    \\
$\Delta$$\delta$ (mas)\tablenotemark{d}  & 0.075 &  0.110     \\
$\chi^2$, $\chi^2_{red}$   &  17.74, 0.77  & 38.02, 1.09 \\
\enddata
\tablenotetext{a}{The parameters presented here were obtained with AGA. Very similar results were obtained with the least-squares fitting method. The astrometric fit includes the orbital motions of a companion. The second column contains the astrometric fit of the new VLBA data. The third column corresponds to the astrometric fit of the combined VLBA data.}
\tablenotetext{b}{Fixed eccentricity.} 
\tablenotetext{c}{Fixed mass of the star.}
\tablenotetext{d}{The rms dispersion of the residual.}
\end{deluxetable*}


%

%

%

%

%

\end{document}